\documentclass{ws-p8-50x6-00}

\usepackage{amsmath,amssymb}

\newcommand{\bsl}{\boldsymbol}

\begin{document}

\title{The doubly heavy baryons in the nonperturbative QCD approach}

\author{I.M.Narodetskii and M.A.Trusov}

\address{ITEP, Moscow, Russia\\ E-mail: naro@heron.itep.ru}

\maketitle

\abstracts{ We present some piloting calculations of the masses of
the doubly heavy baryons in the framework of the simple
approximation within the nonperturbative string approach. The
simple analytical results for dynamical masses of heavy and light
quarks and eigenvalues of the effective QCD Hamiltonian are
presented.}

The purpose of this talk is to present the results of the
calculation \cite{NT01} of the masses and wave functions of the
heavy baryons in a simple approximation within the nonperturbative
QCD (see \cite{Si99} and references therein). The starting point
of the approach is the Feynman-Schwinger representation for the
three quark  Green function in QCD in which the role of the time
parameter along the trajectory of each quark is played by the
Fock-Schwinger proper time. The proper and real times for each
quark related via a new quantity that eventually plays the role of
the dynamical  quark mass. The final result is the derivation of
the Effective Hamiltonian, see Eq. (\ref{EH}) below. In contrast
to the standard approach of the constituent quark model the
dynamical mass $m_i$ is not  a free parameter  but it is expressed
in terms of the current mass $m^{(0)}_i$ defined at the
appropriate scale of $\mu\sim 1$~GeV from the condition of the
minimum of the baryon mass $M_B$ as function of $m_i$:
$\frac{\partial M_B(m_i)}{\partial m_i}=0 $. Technically, this has
been done using the einbein (auxiliary fields) approach, which is
proven to be rather accurate in various calculations for
relativistic systems.

This method was already applied to study baryon Regge trajectories
\cite{FS91} and very recently for computation of magnetic moments
of light baryons \cite{KS00}. The essential point of this talk is
that it is very reasonable that the same method should also hold
for hadrons containing heavy quarks.
As in
\cite{KS00} we take as the universal parameter the QCD string
tension $\sigma$ fixed in experiment by the meson and baryon Regge
slopes. We also include the perturbative Coulomb interaction with
the frozen coupling $\alpha_s(\text{1 GeV})=0.4$.

Consider the ground state baryons without radial and orbital
excitations in which case tensor and spin-orbit forces do not
contribute perturbatively.
5survives in the perturbative approximation.
The EH has the
following form
\begin{equation}
\label{EH}
H=\sum\limits_{i=1}^3\left(\frac{m_i^{(0)2}}{2m_i}+
\frac{m_i}{2}\right)+H_0+V,
\end{equation}
where $H_0$ is the non-relativistic kinetic energy operator,
$m_i^{(0)}$ are the current quark masses and $m_i$ are the
dynamical quark masses to be found from the minimum condition, and
$V$ is the sum of the perturbative one gluon exchange potential
$V_c$ and the string potential $V_{string}$. The string potential
has been calculated in \cite{FS91} as the static energy of the
three heavy quarks: $V_{\text{string}}(\bsl{r}_1,\bsl{r}_2,
\bsl{r}_3)=\sigma R_{\text{min}}$, where $R_{\text{min}}$ is the
sum of the three distances $|\bsl{r}_i|$ from the string junction
point, which for simplicity is chosen as coinciding with the
center--of--mass coordinate.

We use the hyper radial approximation (HRA) in the hyper-spherical
formalism approach. In the HRA the three quark wave function
depends only on the hyper-radius
$R^2=\bsl{\rho}^2+\bsl{\lambda}^2$, where $\bsl{\rho}$ and
$\bsl{\lambda}$ are the three-body Jacobi variables: $
\bsl{\rho}_{ij}=\sqrt{\frac{\mu_{ij}}{\mu}}(\bsl{r}_i-\bsl{r}_j)$,
$ \bsl{\lambda}_{ij}=\sqrt{\frac{\mu_{ij,k}}{\mu}}
\left(\frac{m_i\bsl{r}_i+m_j\bsl{r}_j}{m_i+m_j}-\bsl{r}_k\right)$,
where $\mu_{ij}=\frac{m_im_j}{m_i+m_j}$,
$\mu_{ij,k}=\frac{(m_i+m_j)m_k}{m_i+m_j+m_k}$, and $\mu$ is an
arbitrary parameter with the dimension of mass which drops off in
the final expressions. Introducing the reduced function
$\chi(R)=R^{5/2}\psi(R)$ and averaging $V=V_c+ V_{\text{string}}$
over the six-dimensional sphere one obtains the Schr\"odinger
equation
\begin{equation} \label{shr}
\frac{d^2\chi(R)}{dR^2}+2\mu\left[E_n+\frac{a}{R}-bR-\frac{15}{8\mu
R^2}\right]\chi(R)=0, \end{equation} where
\begin{equation} \label{ab} a=\frac{2\alpha_s}{3}\cdot
\frac{16}{3\pi}\sum\limits_{i<j}\sqrt{\frac{\mu_{ij}}{\mu}},~~~
b=\sigma\cdot\frac{32}{15\pi}\sum\limits_{i<j}\sqrt{\frac{\mu(m_i+m_j)}{m_k(m_1+m_2+m_3)}},
\end{equation}
We use the same parameters as in Ref. \cite{KN00}:
$\sigma=0.17$~GeV, $\alpha_s=0.4$, $m^{(0)}_q=0.009$ GeV,
$m^{(0)}_s=0.17$ GeV, $m^{(0)}_c=1.4$ GeV, and $m^{(0)}_b=4.8$
GeV. We solve Eq. (\ref{shr}) by the variational method
introducing a simple variational Ans\"atz $\chi(R)\sim
R^{5/2}e^{-\mu p^2R^2}$, where $p$ is the variational parameter.
Then the three-quark Hamiltonian admits explicit solutions for the
energy and the ground state eigenfunction:
$E\approx\min\limits_pE(p)$, where
\begin{equation} E(p)=\langle\chi|H|\chi\rangle=
3p^2-(a\sqrt{\mu})\cdot\frac{3}{4}\sqrt{\frac{\pi}{2}}\cdot
p+(b/\sqrt{\mu})\cdot\frac{15}{16}\sqrt{\frac{\pi}{2}}\cdot
p^{-1}. \end{equation}

The dynamical masses $m_i$ and the ground state eigenvalues $E_0$
are given for various baryons in Table 1 of Ref. \cite{NT01}. The
dynamical values of light quark mass
$m_q\sim\sqrt{\sigma}|\sim~450-500$ MeV ($q=u,d,s$) qualitatively
agree with the results of Ref. \cite{KN00} obtained from the
analysis of the heavy--light ground state mesons. For the heavy
quarks ($Q~=~c$ and $b$) the variation in the values of their
dynamical masses $m_Q$ is marginal. This is illustrated by the
simple analytical results for Qud baryons. These results were
obtained from the approximate solution of equation $
\left.\frac{dE}{dp}\right|_{p=p_0}=0$ in the form of expansion in
the small parameters $\xi=\sqrt{\sigma}/m_Q^{(0)}$ and $\alpha_s$.
Omitting the intermediate steps one has
\begin{align*}
E_0&=3\sqrt{\sigma}\left(\frac{6}{\pi}\right)^{1/4}\left(1+A\cdot\xi
-\frac{5}{3}B\cdot\alpha_s+\dots\right)\\
m_q&=\sqrt{\sigma}\left(\frac{6}{\pi}\right)^{1/4}\left(1-A\cdot\xi+B\cdot\alpha_s+
\dots\right),\\
m_Q&=m_Q^{(0)}\left(1+2A\cdot\xi^2+\dots\right)
\end{align*}
where for our variational Anz\"ats
$A=\frac{\sqrt{2}-1}{2}\left(\frac{6}{\pi}\right)^{1/4}$,~
$B=\frac{4+\sqrt{2}}{18}\sqrt{\frac{6}{\pi}}$. Accuracy of this
approximation is illustrated in Table 1.

\begin{table}
\label{mytab} \caption{Comparison of results of  analytical and
numerical variational calculations for $\Lambda_b$ and $\Lambda_c$
baryons (all quantities are in units of GeV)}
\begin{center}
\begin{tabular}{|c|c|c|c|c|}
\hline Baryon & \multicolumn{2}{|c|}{$\Lambda_b$} &
\multicolumn{2}{|c|}{$\Lambda_c$}\\ \hline  & Numerical &
Analytical & Numerical & Analytical \\  & calculation &
calculation & calculation & calculation  \\ \hline $E_0$ & 1.06 &
1.08 & 1.18 & 1.16
\\ \hline $m_q$ & 0.56 & 0.56 & 0.52 & 0.53 \\ \hline $m_Q$ & 4.84
& 4.82 & 1.50 & 1.47 \\ \hline
\end{tabular}
\end{center}
\end{table}

To calculate hadron masses we, as in Ref. \cite{FS91}, first
renormalize the string potential:  $V_{\text{string}}\to
V_{\text{string}}+\sum\limits_iC_i$, where the constants $C_i$
take into account the residual self-energy (RSE) of quarks. In
what follows we adjust the RSE constants $C_i$ to reproduce the
center-of-gravity for baryons with a given flavor. As a result we
obtain $ C_q=0.34,~~~C_s=0.19,~~~C_c\sim C_b\sim 0.$

We keep these parameters fixed to calculate the masses given in
Table 2, namely the spin--averaged masses (computed without the
spin--spin term) of the lowest double heavy baryons. In this Table
we also compare our predictions with the results obtained using
the additive non--relativistic quark model with the power-law
potential \cite{BDGNR94}, relativistic quasipotential quark model
\cite{E97}, the Feynman-Hellmann theorem \cite{LRP95} and with the
predictions obtained in the  approximation of double heavy diquark
\cite{LO99}.

In conclusion, we have employed the general formalism for the
baryons, which is based on nonperturbative QCD and where the only
inputs are $\sigma$, $\alpha_s$ and two additive constants, $C_q$
and $C_s$, the residual self--energies of the light quarks. Using
this formalism we have also performed the calculations of the
spin--averaged masses of baryons with two heavy quarks. One can
see from Table 2 that our predictions are especially close to
those obtained in Ref. \cite{BDGNR94} using a variant of the
power--law potential adjusted to fit ground state baryons.

\begin{table}
\label{tab2} \caption{Masses of baryons containing two heavy
quarks}
\begin{center}
\begin{tabular}{|c|c|c|c|c|c|}
\hline State & present work & Ref. \cite{BDGNR94} & Ref.
\cite{E97} & Ref. \cite{LRP95} & Ref. \cite{LO99}\\
\hline $\Xi\{qcc\}$    & 3.69 & 3.70 & 3.71 & 3.66 & 3.48\\
$\Omega\{scc\}$ & 3.86 & 3.80 & 3.76 & 3.74 & 3.58\\ \hline
$\Xi\{qcb\}$    & 6.96 & 6.99 & 6.95 & 7.04 & 6.82      \\
$\Omega\{scb\}$  & 7.13 & 7.07 & 7.05 & 7.09 & 6.92     \\ \hline
$\Xi\{qbb\}$    & 10.16 & 10.24 & 10.23 & 10.24 & 10.09
\\ $\Omega\{sbb\}$   & 10.34 & 10.30 & 10.32 & 10.37 & 10.19
\\ \hline
\end{tabular}
\end{center}
\end{table}

\section*{Acknowledgements}
This work was supported in part by RFBR grants \#\# 00-02-16363
and 00-15-96786.

\end{document}